# Non-local Ponderomotive Nonlinearity in Plasmonics

Pavel Ginzburg, Alex Hayat, Nikolai Berkovitch and Meir Orenstein

Department of Electrical Engineering, Technion, Haifa 32000, Israel

We analyze an inherent nonlinearity of Surface Plasmon Polaritons at the interface of Fermi-Dirac metal plasma, stemming from the depletion of electron density in high-field intensity regions. The derived optical nonlinear coefficients are comparable to the experimental values for metals. We calculate the dispersion relations for the nonlinear propagation of high-intensity Surface Plasmon Polaritons, predicting a nonlinear induced cutoff and vanishing group velocity.

Plasmonic nano-circuits present promising future solutions for on-chip optical interconnect and have gained considerable interest in recent years [1,2]. The sub-diffraction light confinement by Surface Plasmon Polariton (SPP) waves [3,4] opens possibilities for efficient nonlinear interactions and sensing [5]. Plasmonic structures are used as building blocks for metamaterials with negative index of refraction [6]. Along with theoretical studies, basic planar plasmonic structures were realized experimentally, including waveguides [7] and waveguide-based devices, such as Mach-Zehnder interferometers and optical resonators [8]. In the linear regime, behavior of the SPP on a single metal-dielectric interface is well-known [9,10], while the nonlinear regime of plasmonic waveguides [11] is still not fully explored. In general, metal nonlinearities may stem from several physical effects. In metal nano-particles [12] the most significant contribution to the nonlinear response is attributed to the volume limitation of the motion of the charge carriers [13] , while in bulk metals this effect does not exist [14] . However, bulk metals, which serve as the key medium for SPP guiding, are inherently nonlinear. These nonlinearities may induce dramatic changes in device performance [15] and should be considered in plasmonic circuitry modeling [16] and in the analysis of SPP localization [17]. Previously studied nonlinear effects in bulk metals are related to the saturation of interband transitions [18] that are strongly wavelength-dependent, and to laser induced hot-electron contribution [19] (Fermi smearing) and its related electron-electron scattering rate change [20].

Here we present a non-local metal nonlinearity, originating from a collective ponderomotive interaction of charged particles, where the charge carriers are repelled from the high field intensity region, making the dielectric constant $\varepsilon$ average-intensity dependent. We develop a theoretical model for the ponderomotive nonlinearity and its effect on basic SPP propagation, resulting in intensity-dependent SPP light slowing and guiding cutoff. This ponderomotive nonlinearity stems from the inherent equation of motion of electrons in metals - always present in plasmonics even at telecom wavelengths (~1.5 µm) – far from interband transition related wavelengths [21]. Moreover, ponderomotive nonlinearity is related to charge density in electron plasma and is therefore collective and nonlocal in contrast to interband nonlinearities.

In order to model the dependence of metal dielectric coefficient $\varepsilon_M$ on the field intensity, we calculate the modification of charge carrier density by the nonlocal ponderomotive force given by:

$$\overrightarrow{F_{PM}} = -\frac{1}{m}\left(\frac{e}{\omega}\right)^2 \left(\vec{E}\times(\nabla\times\vec{E}) + \vec{E}\cdot\nabla\vec{E}\right) \quad (1)$$

where $e$ and $m$ are the electron charge and mass, $\omega$ is the light angular frequency, and $E$ is the local field amplitude. For an SPP propagating in the z direction, the resulting force is always in the x direction – normal to the surface. The first term on the right-hand side of Eq. 1, describes the effect of the magnetic field contribution to the Lorentz force related to the longitudinal component of the electric field $E_z$ (parallel to the surface and SPP propagation direction) pulling electrons from the surface ($F_{PM}\hat{x}$), in the regular SPP propagation regime where $E_z > E_x$. The second term on the right-hand side of Eq. 1 describes the effect of the electric field normal to the surface $E_x$, which becomes significant near the SPP cutoff condition resulting in $F_{PM}\hat{x}$ as well. The force can be recast in a simpler form $\overrightarrow{F_{PM}} = -\frac{1}{2m}\left(\frac{e}{\omega}\right)^2 \nabla\left(|E|^2\right)$ and the corresponding ponderomotive potential is:

$$\Phi_{PM} = e^2 |E|^2 \big/ 2m\omega^2 \qquad (2)$$

The introduction of the ponderomotive potential avoids the need for tracking the dynamics of carriers, treating the population statistically.

In contrast to the Maxwell-Boltzmann distribution usually encountered in plasma physics of gases [22,23], in metals the ponderomotive potential modifies the Fermi-Dirac electron distribution of the quasi thermal equilibrium Drude model electron plasma, resulting in the following carrier density:

$$n_{PM} = \frac{1}{3\pi^2}\left(\frac{2m}{\hbar^2}\right)^{3/2} \left(E_f - \Phi_{PM}\right)^{3/2} \qquad (3)$$

where $E_f$ is the Fermi energy assumed to be located deep in the band, such that zero temperature approximation for the distribution is applicable. The resulting intensity-dependent metal dielectric constant is:

$$\varepsilon_{PM}\left(|E|^2\right) = 1 - \frac{e^2}{3\pi^2 \varepsilon_0 m\omega^2}\left(\frac{2m}{\hbar^2}\right)^{3/2}\left(E_f - \frac{e^2|E|^2}{2m\omega^2}\right)^{3/2} \qquad (4)$$

For achievable field intensities the leading term in Taylor series yields a Kerr-like nonlinearity

$$\varepsilon_{PM}\left(|E|^2\right) = 1 - \left(\frac{\omega_p}{\omega}\right)^2 + \frac{3}{2}\left(\frac{\omega_p}{3\pi^2 \varepsilon_0 \hbar m e}\right)^{2/3} \frac{e^4}{\omega^4}|E|^2 = \varepsilon_M + \chi_{PM}|E|^2 \qquad (5)$$

where $\omega_p$ is the plasma frequency for low field intensity, $\varepsilon_0$ the vacuum permittivity, $\varepsilon_M$ the linear part of he metal dielectric constant and $\chi_{PM}$ the nonlinear ponderomotive susceptibility. This Kerr-like coefficient is highly dispersive ($\sim 1/\omega^4$) and for telecom wavelengths is on the order of $\chi_{PM} \sim 10^{-18} \, m^2/V^2$, comparable to that of nonlinear glasses. The relatively high nonlinearity significantly influences SPP propagation and localization phenomena, important in plasmonics. In the following, although it has a nonlocal origin, the nonlinear dielectric constant is taken here as local (k-independent) which is a very good approximation far enough from the plasma frequency.

We study the nonlinear propagation effects due to the ponderomotive potential in the most basic structure supporting SPPs, namely a single metal/dielectric interface, where $z$ is the SPP propagation direction and $x$ is the transverse coordinate. The nonlinear dispersion relations were extracted by several methods. The first is a crude effective index approach, using the nonlinear term for the SPP dispersion [9]:

$$\beta = \sqrt{\frac{\varepsilon_D \varepsilon_{PM}}{\varepsilon_D + \varepsilon_{PM}}} \qquad (6)$$

where $\varepsilon_D$ is the dielectric constant of the substrate, and the nonlinearity stems from substituting the nonlinear $\varepsilon_{PM}$ from Eq. 5. Of course eq. (6) does not represent the accurate nonlinear dispersion.

The second method consistently extracts the nonlinear dispersion relations from the appropriate nonlinear Maxwell's equations and boundary conditions without explicit knowledge of the mode shape [24,25]. We assert here that $E_z >> E_x$, (e.g. for gold the $E_z/E_x$ ratio is about $10$ at $\lambda=1.55\mu m$), such that propagation of a single component $E$ field is considered. The resulting nonlinear dispersion relation is:

$$\frac{\varepsilon_D}{\sqrt{\beta^2 - \varepsilon_D}} + \sqrt{(\beta^2 - \varepsilon_M) - \frac{1}{2}\chi_{PM} E_{z0}^{\ 2} \frac{(\varepsilon_M + \chi_{PM}|E|^2)}{\beta^2 - (\varepsilon_M + \chi_{PM}|E|^2)}} = 0 \qquad (7)$$

where $E_{z0}$ is the magnitude of the electrical field phasor on the interface. This method is expected to be inaccurate near the modal cutoff point – where the magnitude of both $E$ components is comparable (e.g for the linear case near the SPP frequency)

Finally, a full vectorial method was applied [26] (beyond the approximation of $E_z >> E_x$) yielding the accurate dispersion relation as coupled nonlinear equations:

$$\left[\varepsilon_M + \frac{\varepsilon_D^{\ 2}}{\beta^2 - \varepsilon_D}\right] E_{z0}^{\ 2} - \varepsilon_D \beta \frac{E_{z0} E_{x0}}{\sqrt{\beta^2 - \varepsilon_D}} - \frac{1}{2}\chi_{PM} E_{x0}^4 + \frac{1}{2}\chi_{PM} E_{z0}^4 = 0$$

$$\beta \varepsilon_D \frac{E_{z0}}{\sqrt{\beta^2 - \varepsilon_D}} = \left(\varepsilon_M + \chi_{PM} E_{x0}^2 + \chi_{PM} E_{z0}^2\right) E_{x0} \qquad (8)$$

where $\beta$ and $E_{x0}$ are the respective unknown propagation constant and $x$-component of the electrical field phasor amplitude on the interface, while $E_{z0}$ is related to a surface intensity parameter by: $E_{av}^{\ 2} = E_{x0}^2 + E_{z0}^2$.

The propagation constants calculated by the 3 methods mentioned above are depicted on Fig. 1 as a function of the field magnitude ($E_{av}$) on a gold-air interface at a wavelength of λ=1.55μm. All of them exhibit an intensity cutoff where the propagation constants diverge, while the first two approximated methods underestimate the required intensity to reach this cutoff. A descriptive explanation of the mechanism is related to the electron depletion in regions of high field intensity near the metal-air interface, due to the field gradient generated by the field penetration into the metal. Both the transverse and longitudinal components of the field have the same decay depth (for the linear case) into the metal and thus contribute equally (weighted by their respective field magnitudes) to this depletion. As a result of the depletion – the metal dielectric constant in a layer near the interface approaches the critical value of -1 and thus the propagation constant is enhanced (Fig. 1) (easily seen from Eq. (6) $\beta = \sqrt{\varepsilon_{PM}/(1+\varepsilon_{PM})}\Big|_{\varepsilon_{PM} \to -1} \to \infty$). Furthermore, the field penetration into air is reduced while the penetration into the depleted metal is increased and the skin depths become comparable in both sides of the interface. For example, a linear SPP has a penetration depth into metal and air $d_M = \lambda\sqrt{-(\varepsilon_M+1)/\varepsilon_M^2}/2\pi$ and $d_A = \lambda\sqrt{-(1+\varepsilon_M)}/2\pi$ respectively, which are changed from $0.016\lambda$ to $0.08\lambda$ in metal and from $1.6\lambda$ to $0.16\lambda$ in air upon depleting the metal $\varepsilon_M$ from -100 to -2. Combining this mode shrinkage with the reduction of the field magnitude at the interface (reduction of $|\varepsilon_{PM}|/\varepsilon_D$) results in more power being pushed into the metal. Since power propagation in air and metal are opposite in direction – a net reduction of forward power transfer is experienced – directly related to light slowing. The group velocity reduction is evident from the full fledged calculation of the nonlinear dispersion relations (Fig. 2) and explicitly as the group index (Fig. 3). The resulting nonlinear SPP cutoff is reached when the intensity induced mode reshaping results in equal power transport in metal and air.

It should be emphasized that although we presented the exact full vectorial nonlinear dispersion relations, they are given as a function of the $E_{av}$ parameter – namely the field intensity at the metal- air interface rather than using a more intuitive (practical) value of the overall input intensity, since for the latter - we should solve numerically the nonlinear mode field distributions – which is not the purpose of this paper. Due to the discussed shrinkage of the mode at high intensity and the preferred field penetration to the metal, the interface field intensity in the metal is enhanced superlinearly with the input beam intensity, which means that cutoff value is expected at much lower input intensity than may be inferred from the figures.

In conclusion, we have analyzed metal nonlinearity due to the ponderomotive force which repels charge carriers from high field intensity regions and introduced its effect on the metal susceptibility – which can be approximated as a dispersive Kerr like effect – with magnitude similar to that of nonlinear glass. The propagation of a single surface nonlinear SPP was studied and the nonlinear dispersion curve was derived analytically - without the actual numerical solution the modal fields. The cutoff and slow light features of the nonlinear dispersion were explained.

**References :**

[1] W. L. Barnes, A. Dereux, and T. W. Ebbesen, Nature **424**, 824 (2003).

[2] A. Pyayt, B. J. Wiley, Y. Xia, A. Chen, and L. Dalton, Nature Nanotechnology, **3**, 660-665, (2008).

[3] P. Ginzburg, D. Arbel, and M. Orenstein, Opt. Lett. **31**, 3288-3290 (2006).

[4] P. Ginzburg and M. Orenstein, Opt. Express **15**, 6762-6767 (2007).

[5] S. Nie, and S.R. Emory, Science **275**, 1102 (1997).

[6]M. Shalaev, W Cai, U. K. Chettiar, H Yuan, A. K. Sarychev, V. P. Drachev, and A. V. Kildishev, Opt. Lett., **30**, 3356-3358, (2005).

[7] S. I. Bozhvolny, V. S. Volkov, E. Devaux, and W. Ebbesen, Phys. Rev.Lett. **95**, 046802, (2005).

[8] S. I. Bozhevolnyi, V. S. Volkov, E. Devaux, J. Laluet, and T. W. Ebessen, Nature **440**, 508-511 (2006).

[9]H. Raether, Surface Plasmons (Springer-Verlag, 1988), Vol. 111

[10] A. Maier, Plasmonics: Fundamentals and Applications, chapter 2.2, pp.25-30. (Springer Science + Business Media LLC, 2007).

[11] A. R. Davoyan, I. V. Shadrivov, and Y. S. Kivshar, Opt. Express **16**, 21209-21214 (2008).; A. R. Davoyan, I. V. Shadrivov, and Y. S. Kivshar, "Self-focusing and generation of spatial plasmon-polariton solitons", unpublished.

[12] H. Wang, D.W. Brandl, F. Le, P. Nordlander, and, N. J. Halas, Nano Letters **6** (4), 827-832, (2006).

[13] F. Hache, D. Ricard, C. Flytzanis, and U. Kreibig., App. Phys. A, **47**, 347, (1988).

[14] F. Hache, D. Ricard, and C. Flytzanis, J. Opt. Soc. Am. B **3**, 1647-1655 (1986).

[15] S. Palomba and L. Novotny, Phys. Rev. Lett. **101**, 056802 (2008).

[16] G.A. Wurtz, A.V. Zayats, Laser & Photonics Review **2**, 125-135, (2008).

[17] M. I. Stockman, S. V. Faleev, and D. J. Bergman, Phys. Rev. Lett. **87**, 167401-1-4 (2001).

[18] G. Piredda, D. D. Smith, B. Wendling, and R. W. Boyd, J. Opt. Soc. Am. B **25**, 945-950 (2008).

[19] N. N. Lepeshkin, A. Schweinsberg, G. Piredda, R. S. Bennink, and R. W. Boyd, Phys. Rev. Lett. **93**, 123902 (2004).

[20] M. Perner, P. Bost, U. Lemmer, G. von Plessen, J. Feldmann, U. Becker, M. Mennig, M. Schmitt, and H. Schmidt., Phys. Rev. Lett. **78**, 2192 (1997).

[21] K. F. MacDonald, Z. L. Sámson, M. I. Stockman, and N. I. Zheludev, Nature Photon., **3**, 55, (2008).

[22] P. Kaw, G. Schmidt, and T. Wilcox, Phys. Fluids **16**, 1522-1525 (1973).

[23] C. E. Max, Phys. Fluids **19**, 74 (1976).

[24] M. Y. Yu, Phys. Rev. A **28**, 1855 (1983).

[25] K.M.Leung, Phys. Rev. A **31**, 1189. (1985).

[26] D. Mihalache, G. I. Stegeman, C. T. Seaton, E. M. Wright, R. Zanoni, A. D. Boardman, and T. Twardowski, Opt. Lett. **12**, 187-189 (1987).

**Figure captions:**

**Fig. 1.** (Color online) The nonlinear part of the propagation constant $\beta_{NL}$ normalized by the linear part $\beta_L$ vs. the electric field phasor magnitude in the metal at the air-gold interface for wavelength of 1.5μm calculated by: effective index method (dashed line – red), scalar dispersion equation (dash-dot black) and the full vectorial method (solid line - blue) .

**Fig. 2.** (Color online) Nonlinear dispersion relation of a single-surface SPP on air-gold interface at different interface electric field amplitudes: dashed (blue) – 12GV/m, circles (red) – 11.5 GV/m, X's (black) – 11GV/m, diamonds (green) – 10.5GV/m, triangles (brown) – 10GV/m, stars (purple) – 9.5GV/m. The inset is the nonlinear propagation constant normalized by the linear one vs. the wavelength and field amplitude.

**Fig. 3.** (Color online) Group index vs. wavelength at different electric field amplitudes: dashed (blue) – 12GV/m, circles (red) – 11.5 GV/m, X's (black) – 11GV/m, diamonds (green) – 10.5GV/m, triangles (brown) – 10GV/m, stars (purple) – 9.5GV/m. The inset is the nonlinear group index normalized by the linear one vs. the wavelength and field amplitude.

**Figure 1**

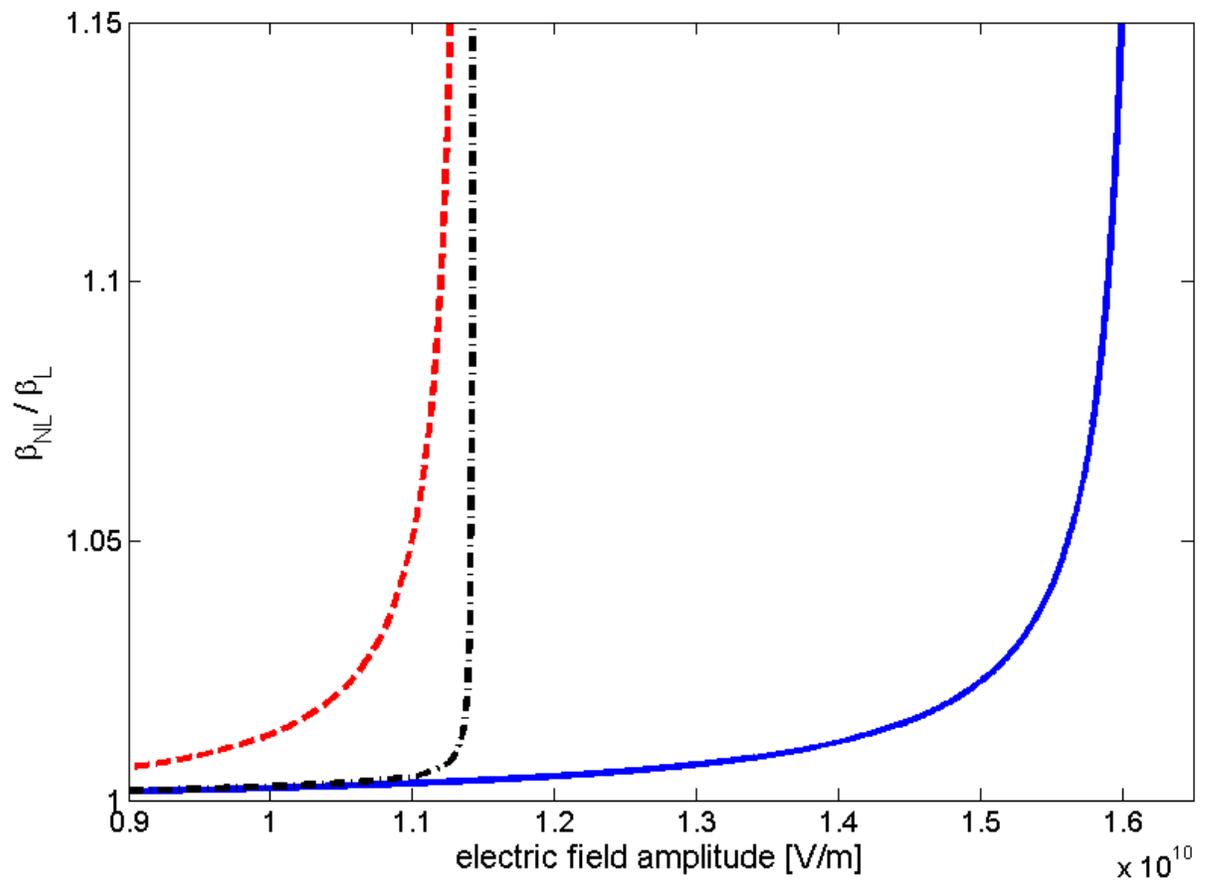

**Figure 2**

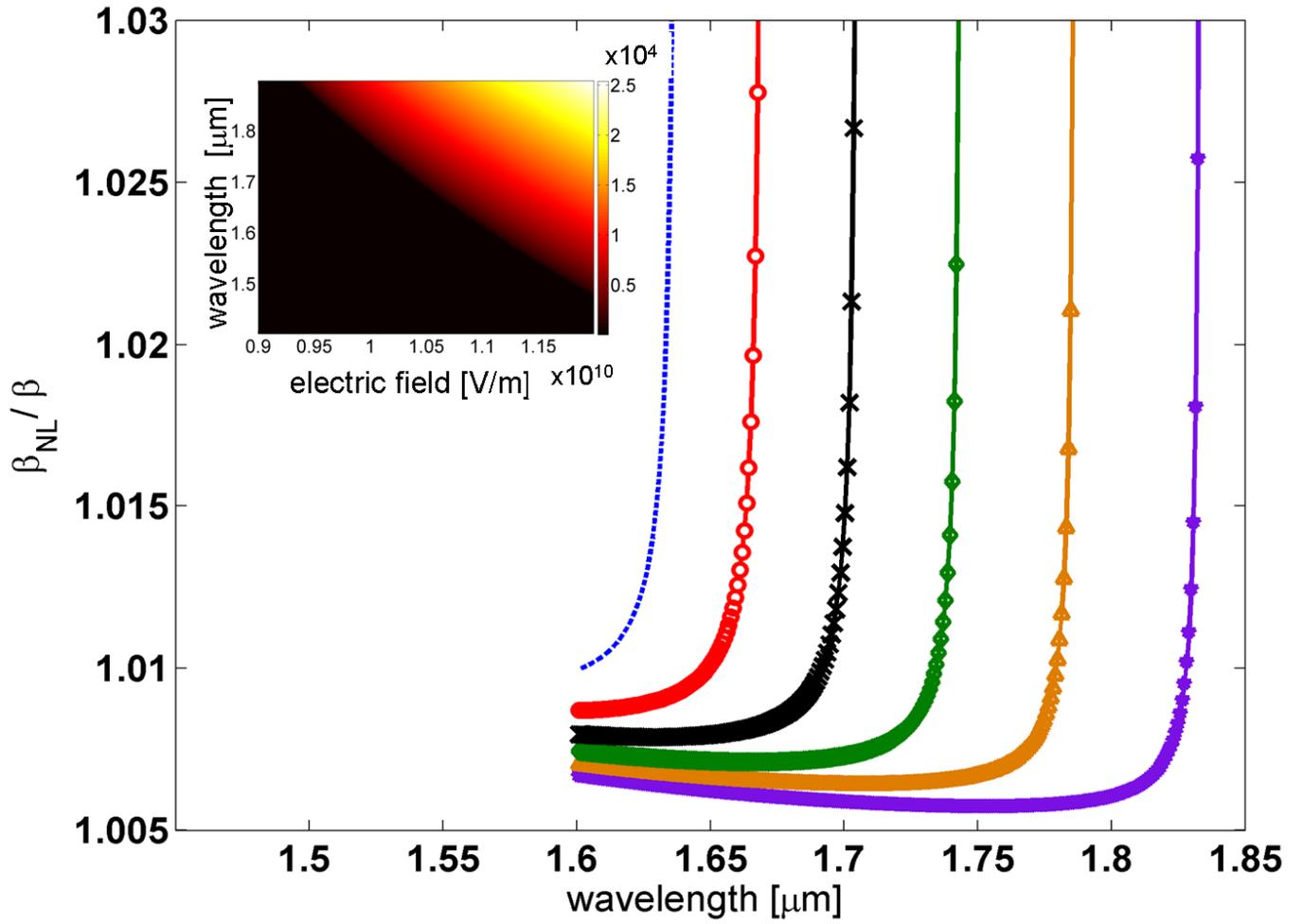

**Figure 3**

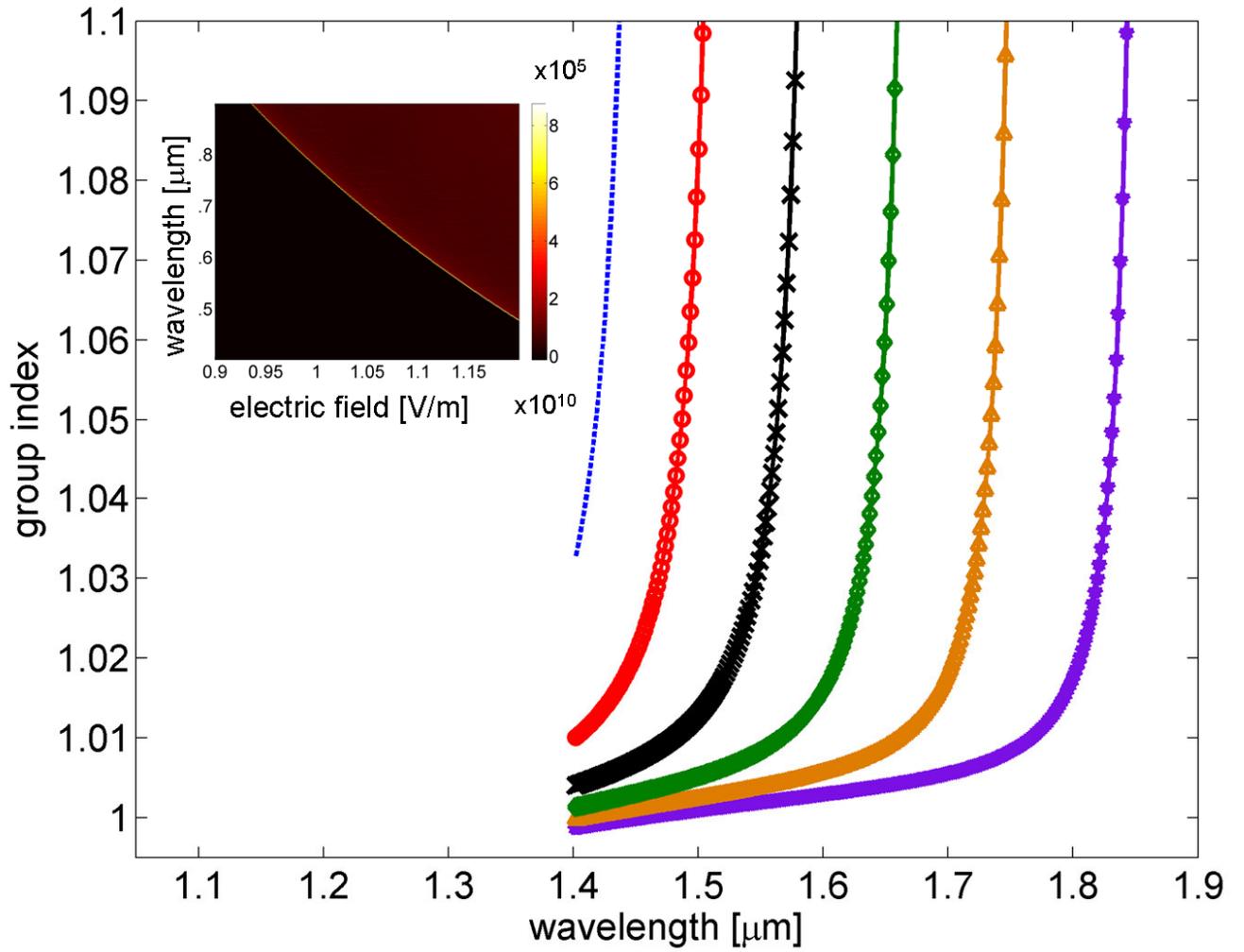